# The CNAO Dose Delivery System for modulated scanning ion beam radiotherapy

S. Giordanengo[a)], M. A. Garella[a,b)], F. Marchetto[a)], F. Bourhaleb[a,c)*], M. Ciocca[b)], A. Mirandola[b)], V. Monaco[a,c)], M. A. Hosseini[a,c)**], C. Peroni[a,c)], R. Sacchi[a,c)], R. Cirio[a,c)] and M. Donetti[a,b)]

a) Istituto Nazionale di Fisica Nucleare, Section of Torino, 10125 Torino, Italy;
b) Centro Nazionale Adroterapia Oncologica, 27100 Pavia, Italy;
c) University of Torino, Physics Department, 10125 Torino, Italy;
*) Now at I-see srl, 10125 Torino, Italy
**) Now at Ionizing and Non-Ionizing Radiation Protection Research Center, Shiraz University of Medical Sciences, Shiraz, Iran.

## Abstract

**Purpose:** This paper describes the system for the dose delivery currently used at the Centro Nazionale di Adroterapia Oncologica (CNAO) for ion beam modulated scanning radiotherapy.

**Methods:** CNAO Foundation, Istituto Nazionale di Fisica Nucleare (INFN) and University of Torino have designed, built and commissioned a Dose Delivery System (DDS) to monitor and guide ion beams accelerated by a dedicated synchrotron and to distribute the dose with a full 3D scanning technique. Protons and carbon ions are provided for a wide range of energies in order to cover a sizeable span of treatment depths.

The target volume, segmented in several layers orthogonally to the beam direction, is irradiated by thousands of pencil beams which must be steered and held to the prescribed positions until the prescribed number of particles has been delivered. For the CNAO beam lines these operations are performed by the DDS. The main components of this system are two independent beam monitoring detectors, called BOX1 and BOX2, interfaced with two control systems performing the tasks of real-time fast and slow control, and connected to the scanning magnets and the beam chopper. As a reaction to any condition leading to a potential hazard, a DDS interlock signal is sent to the Patient Interlock System which immediately stops the irradiation. The essential tasks and operations performed by the DDS are described following the data flow from the Treatment Planning System (TPS) through the end of the treatment delivery.

**Results:** The ability of the DDS to guarantee a safe and accurate treatment was validated during the commissioning phase by means of checks of the charge collection efficiency, gain uniformity of the chambers and 2D dose distribution homogeneity and stability. A high level of reliability and robustness has been proven by three years of system activity needing rarely more than regular maintenance and working with 100% up-time. Four identical and independent DDS devices have been tested showing comparable performances and are presently in use on the CNAO beam lines for clinical activity.

**Conclusions:** The Dose Delivery System described in this paper is one among the few worldwide existing systems to operate ion beam for modulated scanning radiotherapy. At the time of writing it has been used to treat more than 350 patients and it has proven to guide and control the therapeutic pencil beams reaching performances well above clinical requirements. In particular, in terms of dose accuracy and stability, daily Quality Assurance measurements have shown dose deviations always lower than the acceptance threshold of 5% and 2.5% respectively.

## I. INTRODUCTION

Radiation therapy aims to deliver the prescribed amount of dose to the tumor at the same time sparing the surrounding tissues as much as possible. Radiation therapy with charged hadrons has been proposed as an option for patients for whom traditional treatment with photons is not expected to give the desired results. The interaction properties of charged particles with matter leads to a dose deposition highly peaked at end of the range, the Bragg peak, where carbon ions feature an increased radiobiological effectiveness (RBE). These are the main reasons for exploiting charged hadrons in radiotherapy.

Most early ion beam delivery systems [1, 2, 3] provided uniform lateral dose profiles. More recent systems [4, 5, 6, 7] have been designed to provide fluence modulated lateral dose profiles. Using pristine monoenergetic beamlets at well-defined penetration depths, highly conformal dose distributions are achieved without any need of patient-specific hardware. The modulated scanning ion beam technique is currently used only in few centers worldwide [8, 9, 10] and is under development in several existing and proposed facilities [11, 12, 13].

The use of a synchrotron to accelerate ions in a wide range of energies results in a fully three-dimensional dose delivery. The target volume, segmented in several layers orthogonal to the beam direction and corresponding to different beam energies, is irradiated by the superposition of a sequence of beamlets, each delivering a defined number of particles at a specific position. In order to perform such operations with a high precision, a dedicated fast, accurate and redundant Dose Delivery System (DDS) is required. Therefore, by monitoring the particle flux for each beamlet, by requiring the proper beam energy at the beginning of each spill and by controlling the steering magnets to scan the beam laterally, the DDS plays a crucial role within the complex architecture of a facility for modulated scanning ion beam radiotherapy.

This paper describes the DDS in use at Centro Nazionale di Adroterapia Oncologica (CNAO), a hospital-based facility built in Pavia, Italy, to treat cancer using beam scanning techniques and to perform advanced research in the fields of hadrontherapy, radiobiology and physics [14]. The CNAO facility started treating patients with protons in September 2011 and with carbon ions one year later. At the time of writing 350 patients, mainly with chordoma and chondrosarcoma or squamous cell carcinoma located in the head and neck or spine regions, have been treated, 100 with protons and 250 with carbon ions.

In section II an overview of the CNAO treatment system is presented with a brief description of the synchrotron, followed by a summary of the beam delivery technique together with the characteristics of the scanning elements in order to illustrate the role of the DDS components and their interfaces. The technical details of the DDS are reported in section III, while the description of the tasks performed by the system can be found in section IV. Finally, section V is devoted to the treatment safety issues with the description of interlocks handling, while examples of the system performance are reported in the last section (VI).

## II. OVERVIEW OF THE CNAO TREATMENT SYSTEM

### II.A Accelerator and beam characteristics

The CNAO synchrotron [14] has been designed to accelerate protons or carbon ions to energies corresponding to depth in water from 3 to 32 $g/cm^2$ and from 3 to 27 $g/cm^2$, respectively. Two ion sources, a radio frequency quadrupole and a linear accelerator are located inside the 25 m diameter ring of the synchrotron. Particles are accelerated to the required energy by a radio frequency cavity, which performs also the beam stabilization before the extraction. Once the beam has been accelerated to the required energy, the extraction phase starts and particles are steered through the high-energy beam transport line to the selected treatment room.

The period of the synchrotron cycle is typically between 4 and 5 s; the beam is extracted over a period of approximately 1.5 s (spill length) while the remaining time is used to setup the machine for the next spill.

The beam specifications are listed in Table 1. Changes of energy can be performed between spills and the number of delivered particles can be varied between $10^{10}$ protons or $4\times10^8$ carbon ions per spill down to approximately 10% of these values. The transverse beam dimensions in air at the isocenter vary for protons [carbon ions] from 20 mm [8 mm] FWHM at the lowest energy down to 7 mm [4 mm] at the highest energy.

Table 1. Beam specifications

| Maximum field size | $20\times20$ cm$^2$ at the isocenter |
|---|---|
| Beam range for protons (carbon ions) | 3-32 gcm$^{-2}$ (3-27 gcm$^{-2}$) |
| Beam energy range for protons (carbon ions) | 60-250 MeV (120-400 MeV/u) |
| Min-Max beam intensity for protons (carbon ions) | $10^9$-$10^{10}$ ($4\times10^7$-$4\times10^8$) particles/s |
| FWHM in air of spot at the isocenter for protons (carbon ions) | 7-20 mm (4-8 mm) from highest to lowest energies |
| Beam intensities | 100 %, 50 %, 20 %,10% |

Two treatment rooms are equipped with a fixed horizontal beam line, while in the third room both vertical and horizontal fixed lines are available. In the vertical line the beam is steered to the height of 4 m above the isocenter and then deflected to enter vertically the treatment room. Four identical Dose Delivery Systems, one for each beam line, are used at CNAO. A design of the room 2 with the horizontal and vertical beam lines is shown in Figure 1.

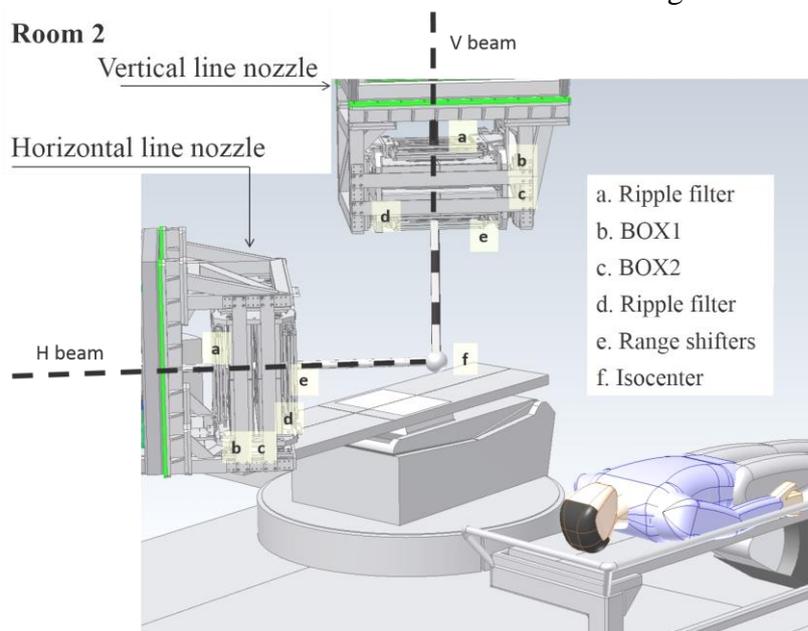

Fig. 1. Room 2 nozzles with horizontal and vertical beam lines: the particles reach the isocenter (f) going through a first ripple filter (a), the beam monitors (b BOX1, c BOX2), a second ripple filter (d), range shifters (e).

As shown in Figure 1, the elements placed downstream of the exit window are the beam monitors and a number of optional beam modifying elements inserted to modify, if necessary, the longitudinal properties of the beam, as described in the following section.

For the horizontal beam lines a pair of dipole magnets, with orthogonally oriented fields, are used to deflect the beam horizontally (X) and vertically (Y), their centers being located at distances of 6.2 m and 5.5 m upstream of the isocenter. A similar arrangement is used in the vertical beam line, with the steering magnets placed at a distance of 2.5 m and 2.0 m upstream of the 90° bending magnet.

### II.B. Beam delivery

The dose delivered using the modulated scanning ion beam technique results from the superposition of a large number of beams conveyed to the target. The irradiated volume is divided into several layers, called slices, orthogonal to the particle propagation direction [7]. Each layer is irradiated by several overlapping iso-energetic beams arranged in a grid of 2-3 mm pitch, called spots, each delivering a dose in small volumes and characterized by the number of particles and by the position in the transverse plane. More than 15-20 daily fractions compose a typical treatment. Each dose fraction is delivered through two or more fields [15, 16] and each field consists of 30 to 100 slices and of 10000 to 50000 spots to cover the tumour volume.

Range shifters can be added to degrade the beam energy for shallow tumour treatments or for finer range shift [17]. Furthermore, for carbon ion beams it has to be emphasized that the Bragg peak is quite narrow, and a uniform longitudinal dose distribution would require the superposition of a large number of energies leading to an inefficient use of the beam. This issue is solved by inserting up to two ripple filters to broaden the pristine Bragg peak [18, 19]. During the treatment, the DDS guides the beam delivery irradiating the spots according to a sequence defined by the Treatment Planning System (TPS). The number of particles delivered to each spot is measured by the beam monitors chambers; when the prescribed number of particles is reached, the beam is steered to the next spot acting on the scanning magnets. A tiny fraction of the dose is therefore delivered between spots and the DDS includes this transient dose in the arrival spot dose evaluation to prevent extra-dose. However, when distance between spots centers is larger than a pre-defined threshold, the irradiation is paused. This procedure is repeated until the last spot of the slice has been irradiated.

The DDS could require several identical spills to deliver the prescribed dose to a given slice each characterized by a defined beam energy. The energy layers versus spills are presented in the diagram of Figure 2, which shows a typical time structure of the CNAO treatment delivery. By optimizing the number of particles in each spot [20], the dose to the target volume can be precisely conformed to the prescription.

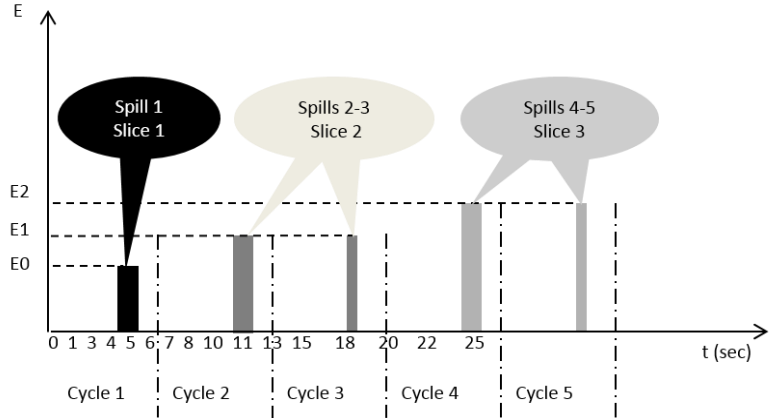

Fig. 2. Diagram to explain the energy layers as a function of time. The spill time structure is evidenced.

### II.C Magnetic scanning system

One of the key elements of the modulated beam scanning technique is the system of steering magnets and power supplies [5, 9]. The details of construction and performances have been reported in [21] and references therein; here we give only a short summary focussing on the aspects which are more relevant to the DDS and the related clinical requirements.

The CNAO scanning magnets provide a $20 \times 20$ cm$^2$ maximum clinical irradiation field corresponding to a maximum bending angle of 21 mrad. Particular care was taken in the design to achieve a uniform magnetic field over the full $12 \times 12$ cm$^2$ aperture area in addition to a low magnetic hysteresis. The power supplies feature a current sensitivity of 55 mA, which corresponds for the lowest beam energies to a 70 and 200 µm position sensitivity at the isocenter for carbon and proton beam respectively.

The other important design performance is the beam steering speed because the beam is not stopped when moving from one position to the next. The magnetic scanning system was carefully designed in order to minimize the transit time between spots which was measured to be less than 200 µs [21], corresponding to an average scanning speed exceeding 15 m/s. The relative effect on the dose depends on the time spent on each spot and on the beam intensity. However, as said before, the transient dose does not imply additional dose to the patient since it is accounted for by the DDS as part of the dose delivered to the arrival spot.

### III. THE CNAO DOSE DELIVERY SYSTEM

After a brief introduction on the roles of the DDS, a detailed description of the system is presented in three sections: components, tasks and operations, interlock management.

### III.A. Introduction

The role of the DDS consists of delivering the dose according to the prescriptions, i.e. irradiating each spot with defined properties: number of particles, selected energy and specified position. The delivery needs to be performed ensuring the safety to the patient.

The above mentioned spot properties are provided as a DICOM RT file by the Syngo Siemens version VC10 TPS, which supports the delivery of both protons and carbon ions with scanning modality.

The patient related pre-treatment data flow starting from the TPS through the synchrotron control and to the DDS is shown in Figure 3.

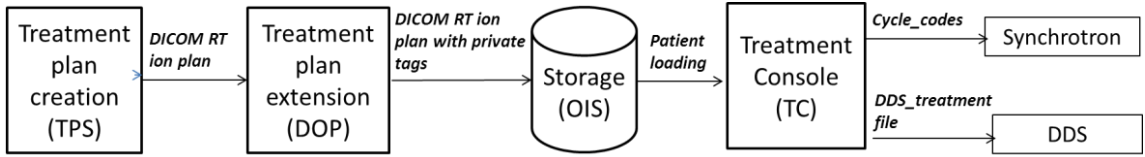

Fig. 3. Pre-treatment data flow starting from the TPS down to the synchrotron control and the DDS through the Dicom Operating Powering (DOP), the Oncology Information System (OIS) and the Treatment Console (TC).

The DICOM Operating Powering (DOP) software tool adds information, stored in private DICOM tags, to the TPS plan in order to completely define the accelerator settings and to optimize the delivery. Among these, we mention optimization of the delivery path, instruction to switch-off the irradiation when large spot-to-spot distance needs to be covered, selection of beam intensity and use of repainting techniques to mitigate the effects of target motion. The extended treatment plan is stored in the Oncology Information System (OIS), Elekta Mosaiq version 2.50. A few minutes before the treatment starts, the Treatment Console (TC) receives the patient plan from OIS and sends to the accelerator control a list of binary codes (cycle-codes) defining the energy, beam size, flux and particle type for each spill. In parallel the *DDS_treatment* file, containing the full list of spots grouped in slices (see Sec II.B), is produced and downloaded. The spot sequence defined in the *DDS_treatment* file is used by DDS during beam delivery to guide the beam during irradiation.

### III.B. Dose Delivery System (DDS) components

The DDS components can be divided in two groups:
- beam detectors placed at the nozzle;
- control system.

### III.B.1. Beam detectors

The DDS measures at a fixed frequency the number of delivered particles, the beam transversal position and dimension by means of five parallel plate ionization chambers filled with nitrogen. The details of construction and performances are reported in [22] while a brief description is given in the following. The monitor chambers are enclosed in 2 independent steel boxes: BOX1 and BOX2 (Figure 4). The BOX1 contains an integral chamber (INT1) with a large area anode for the measurement of the beam flux, integrated over the irradiation area, followed by two chambers with the anodes segmented in 128 strips, 1.65 mm wide, respectively with vertical (StripX) and horizontal (StripY) orientations, which provide the measurement of the beam position and beam width along two orthogonal directions. The BOX2 contains a back-up integral chamber (INT2) followed by a chamber with the anode segmented in $32 \times 32$ pixels, 6.6 mm wide, (PIX). The measurements accomplished by the BOX2 detectors are beam fluence, position and width.

The total water equivalent thickness of these chambers has been measured to be approximately 0.9 mm. The material budget interposed by the beam detectors is spread along approximately 20 cm and it is one of the main contributions to the beam lateral dispersion. To minimize the effect, the boxes are installed closed to the patient, at approximately 70 cm from the isocenter.

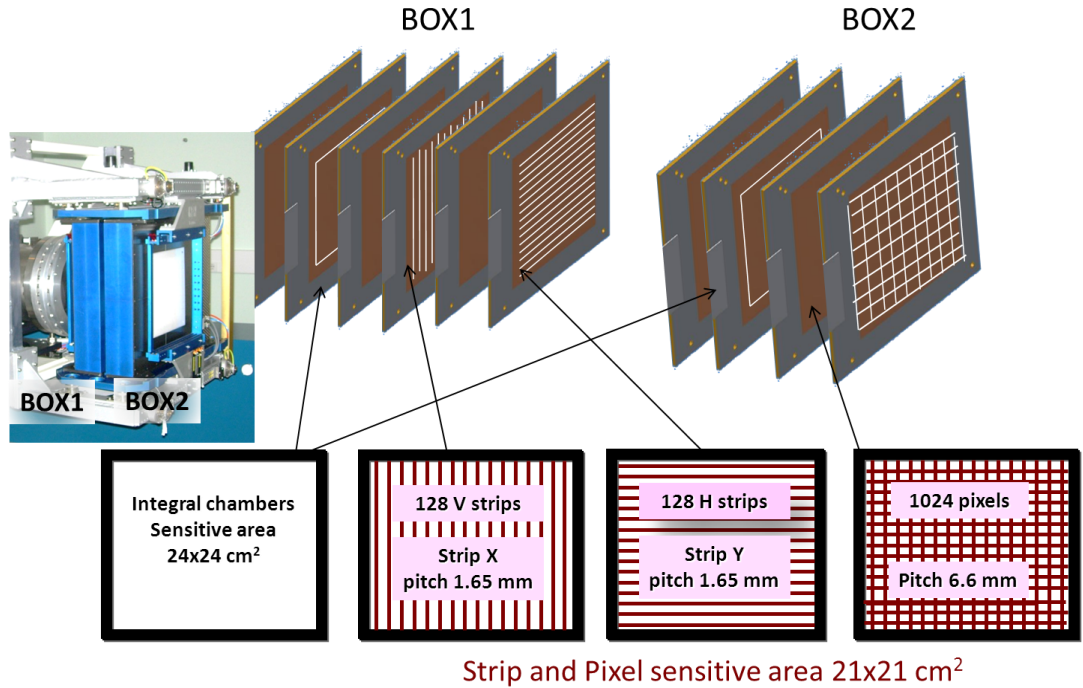

Fig. 4. Detail of the electrode segmentation for beam detectors (BOX1 and BOX2).

The detector front-end readout is based on custom designed boards, which host Application Specific Integrated Circuit electronics (ASIC) custom designed for this purpose. The chosen architecture and technology allow good sensitivity, with a minimum measured charge of 200 fC corresponding to a number of protons ranging from $7.2 \times 10^3$ at the energy of 62 MeV to $1.9 \times 10^4$ at the energy of 226 MeV; similarly, the range of the number of carbon ions extents from 341 at the energy of 115 MeV/u to 767 at 399 MeV/u. The background current is limited to 200 fA; this ensures negligible error on the dose delivered to each spot.

The customized electronics is based on large-scale integration to provide many readout channels with uniform channel-to-channel behavior, and allow a large segmentation of the active area.

The charge collected in the ionization chambers depends on the pressure and temperature of the gas, as well as on the high voltage across the gap. These quantities are monitored by a set of transducers installed in each box and controlled by a Peripheral Interface Controller (PIC). Values are periodically read and checked by the PIC against pre-set values. The measured deviations are used to correct the gain of the chambers. Appropriate interlock procedures are activated whenever any of the values is outside the expected range (see Sec. III.B.2.2).

### III.B.2. DDS control system

The electronics for the control of the whole DDS system (including the detector power supplies, the gas distribution, the interface with the interlock system and the dose delivered recovery system) is hosted in the cabinet shown in Figure 5. To avoid radiation damage, the cabinet is positioned approximately two meters aside of the beam pipe, in a vault located just before the treatment room and protected by a concrete wall.

The architecture of the control system is based on commercial National Instruments (NI) crates and boards. Two Peripheral Component Interconnect (PCI) eXtensions for Instrumentation (PXI) crates are used: the Fast Control (FC), dedicated to the time critical operations performed during the irradiation and the Slow Control (SC) to check for the proper beam modifiers and for the system conditions before each treatment.

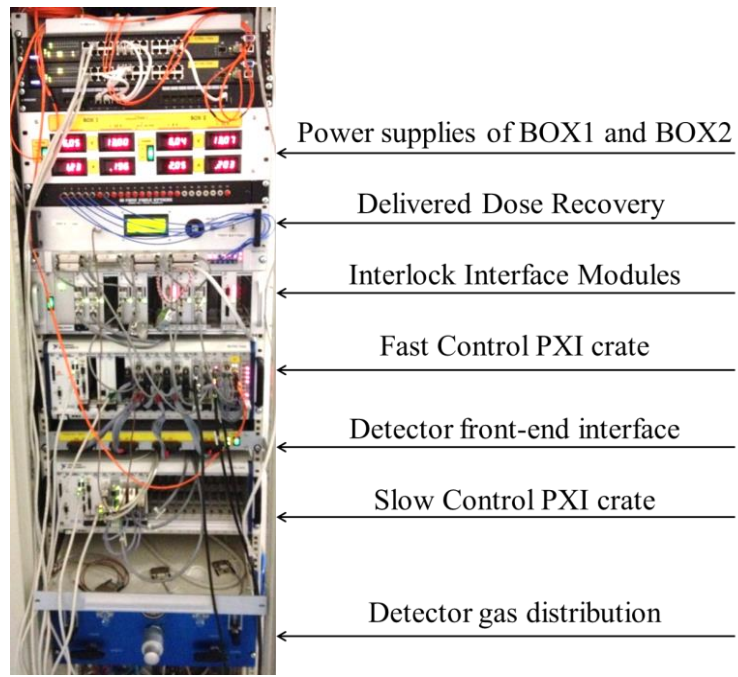

Fig. 5. Cabinet hosting the DDS.

### III.B.2.1. Fast Control

The Fast Control crate includes a CPU running a Real-Time Operating System (RTOS), five NI PXI modules with Field Programmable Gate Array (FPGA) on board, four memory cards, one additional network card and three custom PXI modules.

The main intelligence and the safety of the DDS are entrusted into the 5 FPGAs[1] which work in standalone mode to ensure maximum speed and reliability during the beam extraction phase. While treating a spot, the leading role is played by one of the FPGAs (*Int FPGA*), which drives the delivery by sequentially accessing the prescribed spot properties, stored in the memory boards, and measuring the number of particles delivered through the integral ionization chambers (INT1 and INT2) at a sampling rate of 1 Msample/s. In parallel, the *Strip FPGA* and *Pixel FPGA* measure position and size of the pencil beam in the transversal plane through StripX, StripY and PIX detectors at a frequency of 20 kHz and 10 kHz respectively. A center of gravity algorithm is applied [22] as soon as a pre-defined minimum number of counts is reached in order to reconstruct the beam position and compare the measured value with the planned one. The difference computed by the *Strip* FPGA is used by the *Scan* FPGA to provide a feedback to the scanning magnets, which correct for small position deviations. When the position deviations are larger than a threshold, presently set at 2 mm, the *Strip* FPGA pauses the irradiation acting on the beam chopper [14].

If the number of counts is below the minimum threshold described above, no feedback is applied and the position of the strip with largest number of counts is returned.

The *Scan FPGA* is used to control the scanning magnets by sending the required current to the power supplies and reading back the measured currents at a rate of 40 kHz. In addition, the

---
[1] Three NI 7813R boards and two NI 7811R by National Instruments

*Scan* FGPA has also in charge to switch off the beam when the distance between spot centers is greater than a threshold, (at present 20 mm), by acting on the chopper.

Finally, the *Timing FPGA* interfaces the FC with the Master Timing Generator (MTG) to synchronize the FC operations with the accelerator operations (magnet preparation, beam acceleration, beam extraction, etc...). The MTG broadcasts cycle-codes (list of binary codes, see section III.A) and synchrotron events to several systems and equipment, (e. g. DDS) which need trigger signals from accelerator operations. Most of the other systems read and use timing events, while DDS FC also generates and sends events back to MTG itself. In particular, the FC at the end of each spill sends the events *Repeat_cycle* or *Next_cycle* to ask for a new spill; this could be a repetition of the previous spill in case a slice has not been completed (see section III.C). Additionally the FC by means of the *Timing FPGA* grants the start of spill by replying to the MTG the acknowledge of the received cycle-code.

### III.B.2.2. Slow Control

The measurement of gas temperature and pressure inside the detectors and the control of the high voltage applied to the chambers are important for accurate beam flux measurements. These controls are performed by the Slow Control (SC) system, which also checks for the insertion of the proper beam modifier (section II.B). The SC crate includes one real-time CPU, three digital input-output cards, one additional network adapter, a serial card connected with the internal Slow Control boards of BOX1 and BOX2 [22], and a PXI module to provide signals to the interlock system.

As described in Sec. II.B, range shifters and ripple filters may be needed for specific treatments. These elements are manually mounted and the SC at a frequency of approximately 1 Hz continuously performs the setup verification. Before the beginning of each treatment, if the expected elements are correctly installed and the detector parameters are within the expected range, the SC communicates to the Supervision System (SS) the grant for delivery.

### III.C DDS tasks and operations

In the DDS several tasks are running in parallel, each performing different operations. The list of tasks, subdivided in six main groups, is shown in Table 2.

Tab. 2. List of the major DDS tasks subdivided in 6 groups (columns).

| DDS task groups | | | | | |
|---|---|---|---|---|---|
| **Input data management** | **Treatment delivery** | **Output data management** | **Interlocks management** | **Data publishing** | **Detector management** |
| Beam modifier insertion check | Synchronization with the accelerator | Treatment *Log* file creation | Interlock signal generation | Data publishing to Supervision System | Gas pressure and temperature check |
| Treatment file transformation | Beam monitoring and control | *Log* file to OIS | | Data publishing in the local control room | High voltage control |

| FPGA data creation | Control of the Scanning power supplies | DDS data remote storage | | | |
|---|---|---|---|---|---|
| | DD recovery update | | | | |

The sequence of a treatment can be divided into the four phases indicated in the flowchart of Figure 6. Following this division, the operations performed by the tasks will be described in detail.

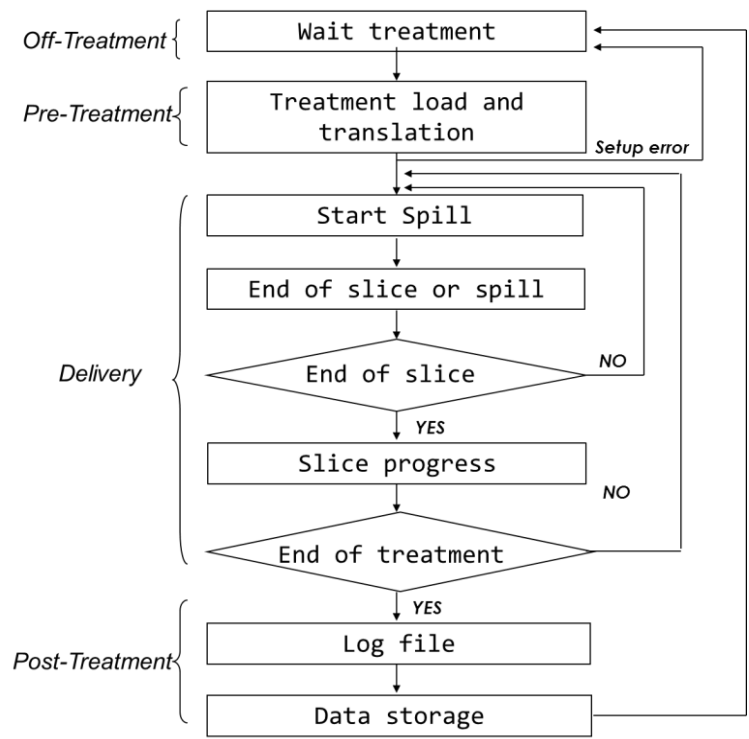

Fig. 6 Flow chart with the DDS operations.

1) Off treatment.

To keep the patient safety to high standards, a large number of system components is continuously checked before treatment delivery. Interlock management, data publishing and detector management are always in operation to issue a warning if any anomalous condition is detected. For example, the *Scan FPGA* sends a pre-configured set of currents to the power supplies of the scanning magnets and receives the measured currents and the power supplies status. Furthermore *Int* FPGA, *Strip* FPGA and *Pixel FPGA* continuously read the detectors in order to monitor the background.

2) Pre treatment.

A few minutes before the treatment, during the procedure for the patient positioning, the Fast Control receives the *DD_Treatment* file via the Treatment Console (TC), as shown in Figure 7. For each spot, the FC performs the conversion of the properties defined by the TPS to the corresponding quantities used internally by the DDS. In detail, the number of particles to be delivered to each spot is converted to number of counts read by the integral chambers, the conversion factor being determined daily with the method described in [23]: the spot coordinates at the iso-center are converted to beam position in strip units at each strip detector. The gas temperature and pressure measured by the SC are accounted for the first conversion. Furthermore, the steering magnet currents, corresponding to the required coordinates, are computed in the pre-treatment phase. Spots to be treated with the same energy are grouped in binary files (*Input FPGA* in Figure 7, named *slice_data*, *sld* file), one for each slice. Later during the treatment, data stored in the *sld* files will be sequentially uploaded into the memory boards, spill by spill, to be accessed by the FPGAs of the FC.

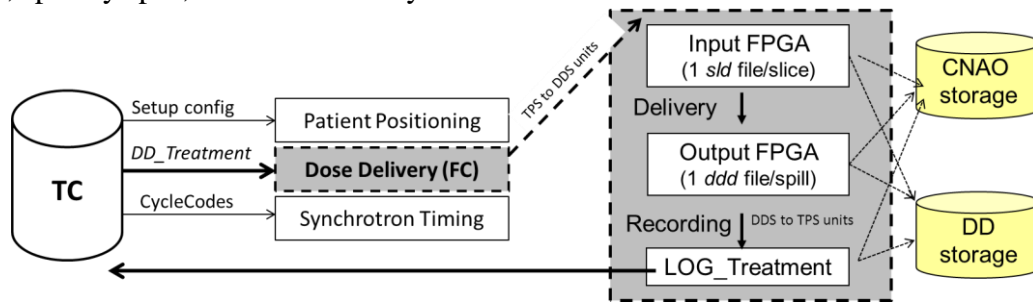

Fig. 7. Data flow during a treatment delivery

The detector background, high voltage, gas temperature and pressure, passive elements setup and interlock registers are also checked; if no problem is found, DD moves into a *ready* status, which will be acknowledged by the SS in order to deliver the beam. In case of need, the SS can require the DDS to abort the treatment procedure just before irradiation, even if the patient is ready and the treatment loaded.

   3) Beam delivery.

Each spill starts after the FC acknowledge of the cycle-code (section III.B.2.1). The MTG triggers the operations to steer the requested beam into the correct treatment room. The delivery starts, following the spot sequence defined in the *sld* files. As soon as one of the two integral chambers reaches the prescribed number of counts for a given spot, a trigger from the *Int FPGA* is sent to the *Scan FPGA* which sets the new currents of the scanning magnets. In parallel the same trigger is received by the *Strip* and *Pixel FPGAs* to load the reference data of the next spot used for the interlock and position feedback. Once the last spot of a slice has been treated (this could take more than one spill as described in Sec. II.B and in Figure 2b) the *Int FPGA* immediately stops the beam delivery by acting on the beam chopper.

At the end of each spill the FC, by means of the *Timing FPGA,* sends a request for a new spill by specifying a cycle-code which can coincide or not with the previous one, depending if the slice has been concluded or not. In the latter case, the FC reloads into the memories the data of the current slice, where the spots already delivered during the previous spill have been removed. As shown in Figure 7, FPGA output data, containing the on-line DDS measurements, are temporary stored in the FC controller as a spill-specific binary file called *dose_delivery_data* (*ddd*). The availability of these files, which contain complete beam monitor readout, is mandatory for the creation of the *Log* file.

   4) Post Treatment.

When the last spot of the last slice is irradiated, a *Log* file is prepared and sent to the OIS, which applies further checks against the planned treatment, in order to grant the end of the dose delivery. A *Log* file is created by FC by using the data stored in the *ddd* files and applying conversions to provide the quantities listed in Tab. 3. The parameters reported in the *Log* file have been chosen on the base of DICOM and OIS requirements. Width and height of the spot profile are not DICOM tags and consequently are not recorded in the *Log* file. At the time of writing, the real-time check of the spot width in term of FWHM is not yet performed by the DDS. However the beam FWHMs are measured during the daily QA tests and compared with the values measured during the TPS commissioning. The FWHMs currently reported in the *Log* file among the slice characteristics, see Tab. 3, are those that have been used by the TPS for the treatment computation.

Finally, the DDS ends the treatment saving *sld* and *ddd* files in a remote storage for off-line data analysis. Such analysis yield important feedback on the performance of the system as described in [24] and [25].

Tab. 3. Main parameters recorded in the *Log* file.

| General | For each slice | For each spot |
|---|---|---|
| - Number of delivered slices<br>- Number of delivered spots<br>- Number of delivered particles<br>- Passive element configuration | - Accelerator cycle-code<br>- Number of delivered spots<br>- Measured cumulative particles<br>- FWHMx, FWHMy (values used by TPS) | - Number of measured particles<br>- Measured spot position (X in mm at isocenter)<br>- Measured spot position (Y in mm at isocenter)<br>- ΔX, ΔY (difference in mm between measured and required spot position) |

### III.C.1 Interlock management

The safety of the treatment mainly relies on two interlock systems: Patient Interlock System (PIS) and Safety Interlock System. These systems collect any detected anomalous or error conditions and either force the immediate interruption of the beam delivery or inhibit the operations as long as the conditions persist.

The PIS is dedicated to the patient safety by acting on the beam chopper to interrupt the treatment when an interlock occurs. The delivery can thus be interrupted for a short time (few seconds) allowing the recovery of the normal treatment execution or can be terminated when the faulty condition causing the interlock persists also after the reset performed by the operator. In the last case, the treatment is interrupted through a safe recovery procedure where the undelivered dose is calculated by the OIS based on the number of particles and spots stored in the DD *Log* file and used to provide a new field to be delivered as soon as possible to complete the treatment fraction.

The DDS error conditions are collected and sent to the PIS by means of hardware signal in order to guarantee a fast reaction. Different Interlocks are generated by the FPGAs of FC, SC when the quantities listed in Tables 4 and 5 are outside the tolerance window. In addition the synchronization between the DDS and the synchrotron is monitored continuously by checking the cycle-code broadcasted by the timing system.

Tab. 4. List of the main Fast Control interlocks with tolerance intervals in use. QInt1 and QInt2 are the number of counts measured by the integral chambers Int1 and Int2 respectively; each count corresponds to 200 fC.

| Critical condition (FC) | Tolerance intervals |
|---|---|
| \|QInt1-QInt2\| | < 100 counts |
| \|QInt1-QInt2\|/QInt1 | < 10% |
| Beam intensity (protons) | $< 3 \times 10^{10}$ protons/s |
| Beam intensity (C ions) | $< 5 \times 10^{8}$ C ions/s |
| Spot position deviation in X | < 2 mm |
| Spot position deviation in Y | < 2 mm |

Tab. 5. List of the main Slow Control interlocks with the tolerance intervals in use for BOX1 and BOX2;

| Critical condition (SC) | Tolerance intervals |
|---|---|
| Gas temperature (BOX1 and BOX2) | 16ºC < T < 26ºC |
| Gas pressure (BOX1 and BOX2) | $9 \times 10^{4}$ Pa < P < $11 \times 10^{4}$ Pa |
| HV (BOX1 and BOX2) | 395 V < HV < 405 V |

One battery backed-up device, called Dose Delivered Recovery system, continuously receives, stores and displays information related to the last treated slice and to the spot of each slice during irradiation. It can work for at least 20 minutes without external power supply and it retains important information in case of major system fault occurring during a treatment.

### III.C.2 Data publishing

Each treatment room is provided with its own local control room, where all the remote operations related to the treatment execution are managed. A dedicated program receives from the FC and displays the measured data, system status and treatment progress. The beam position, the number of monitor counts and the percentage of delivered spots and slices delivered are continuously updated in the DDS monitor display. If any interlock occurs, the display will show it allowing to quickly pinpoint the problem which stops the beam. A snapshot of the DDS monitor display during a carbon ion treatment is shown in Figure 8.

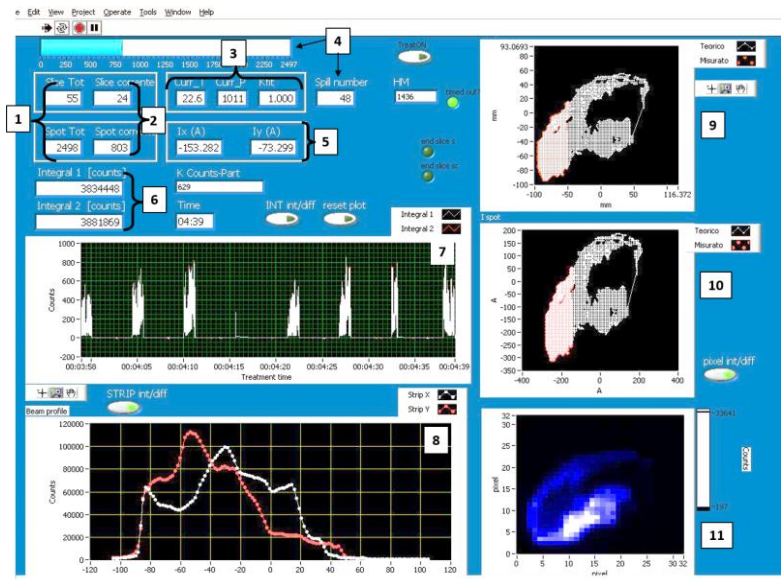

Fig. 8. Snapshot of the data published continuously in the DDS monitor display: total number of slices and spots (1) of this field, slice and spot numbers under delivery (2); measured temperature, pressure and the flux correction factor (3); number of the spill and delivery progress bar (4); PS current set-points (5); INT1 and INT2 total counts (6) flux measured by INT1 and INT2 (7); StripX and StripY total counts (8); spot positions in millimetres (9) and PS currents in ampere (10) measured (red) and required (white); 2D flux measured by PIX chamber (11).

The status of the DDS during the *Off, Pre* and *Post* treatment phases is also monitored by the Supervision System (SS) and displayed in the main control room.

Figure 9 summarizes all the described interfaces used by the FC and SC to perform the DDS tasks together with the type of data exchanged.

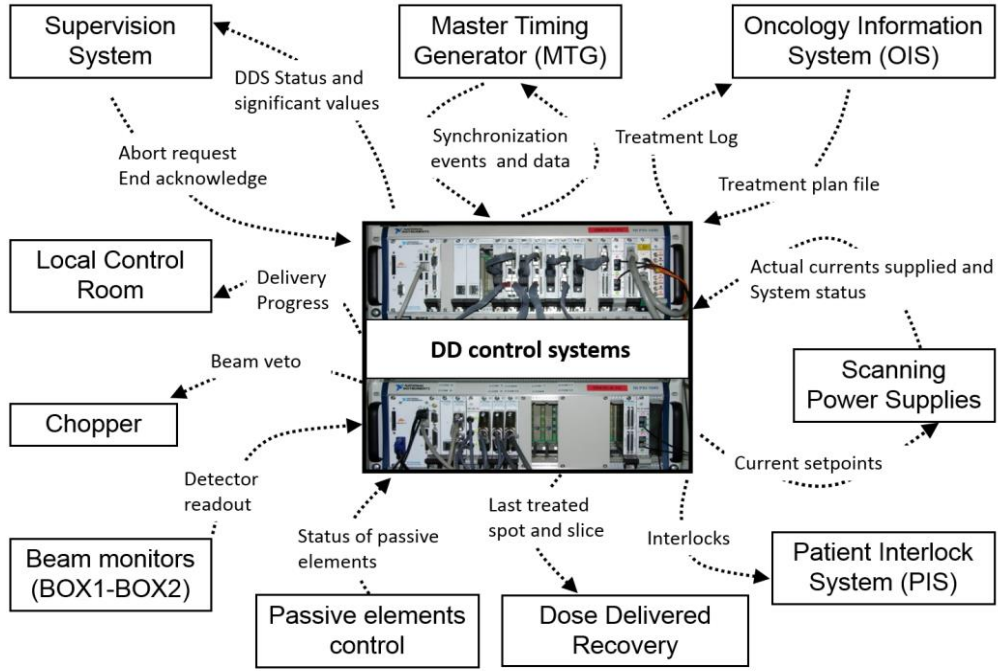

Fig. 9. External systems and devices communicating with the Fast and the Slow Controllers. The exchanged data are described with an arrow indicating the direction of the flow.

## IV. ACCEPTANCE TESTS

The DDS acceptance tests have been performed at CNAO during the pre-clinical beam commissioning phase and the results were found to satisfy the requirements. Although detailed description of the accuracy and the performance of the system are out of the scope of this work and will be reported in a separate paper, a few examples are presented here to show the main characteristics of the DDS: in particular results are shown for the collection efficiency and gain uniformity of the integral chambers and for a uniform square dose distribution delivered at the isocenter plane.

### IV. A. Charge collection efficiency

The integral chamber's efficiency is regularly checked to verify the stability of the collected charge under identical irradiation conditions with both protons and carbon ions. Figure 10 shows examples of efficiency measurements, where the reciprocal of the observed ionization currents (1/counts measured by INT1 and INT2) are shown as a function of the applied high voltage [26].

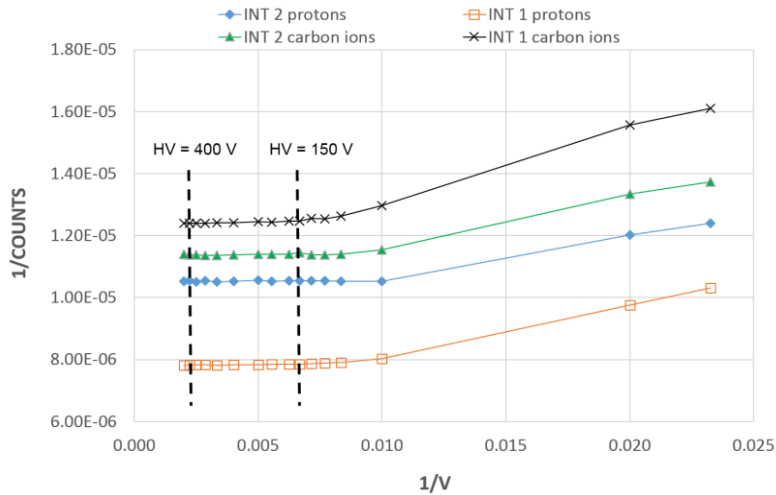

Fig. 10. The reciprocal of the number of counts measured by INT1 and INT2 as a function of the reciprocal applied voltage; the measurements cover the range 50 to 500 V and points corresponding to a high voltage of 150 and 400 V are indicated. The curves were derived using different irradiation conditions and do not overlap.

The results show that at a voltage exceeding 150 V the efficiency saturates and at the nominal operating voltage of 400 V 99.99 % of the saturation value is achieved.

### IV.B. Gain uniformity of the integral chambers

In order to test the 2D gain uniformity of the integral chambers, a $7 \times 7$ square grid of spots, spaced 26 mm from each other, was delivered. The charge collected by an external ionization chamber (PTW T34080) put in air at the isocenter, moved spot by spot synchronously with the beam, was used as a reference to measure the relative gain in each point. The gain was found to be uniform within ± 1.5 %, for all the chambers in use. A Lagrange interpolation on a 1 mm grid was used to obtain a $200 \times 200$ mm$^2$ gain distribution grid shown in Figure 11 which is used by the DDS to apply fine non-uniformity corrections to the monitor counts.

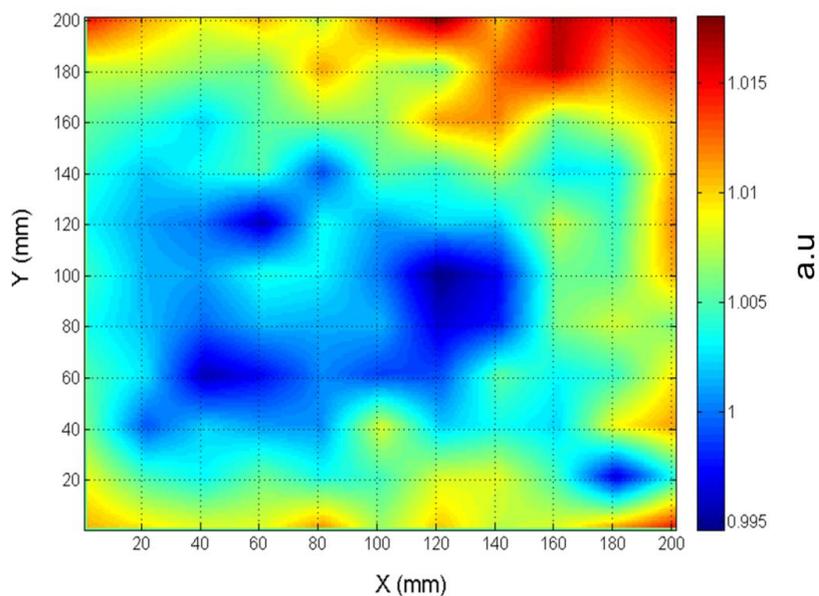

Fig. 11 Example of 2D gain distribution of an integral chamber.

### IV.C. Dose Distributions

2D uniform fields are periodically delivered for quality assurance tests by irradiating a regular grid of spots in air at the isocenter, each spot with the same number of particles. Square fields of $150 \times 150$ mm$^2$ and $60 \times 60$ mm$^2$ size with spot spacing of 3 mm (protons) and 2 mm (carbon ions) were analyzed for different beam lines, particles and energies. The dose uniformity of the delivered fields is measured by using EBT3 radiochromic films [27]. The uniformity is evaluated in terms of flatness, (Imax-Imin)/(Imax+Imin) within the 80% of the field size (70% for diagonals) and results from different beam lines, particles per spot, particle type and energies are found to be within the tolerance in use at CNAO of 5%. Figure 12 shows measured field profiles along main axes (x and y) and diagonals for a homogeneous $60 \times 60$ mm$^2$ square field irradiated following a $30 \times 30$ grid of 2 mm spaced spots; $5 \times 10^4$ carbon ions per spot with energy of 221.45 MeV/u and 6 mm FWHM were delivered in air at the isocenter. The measured flatness for that field is 2.5%.

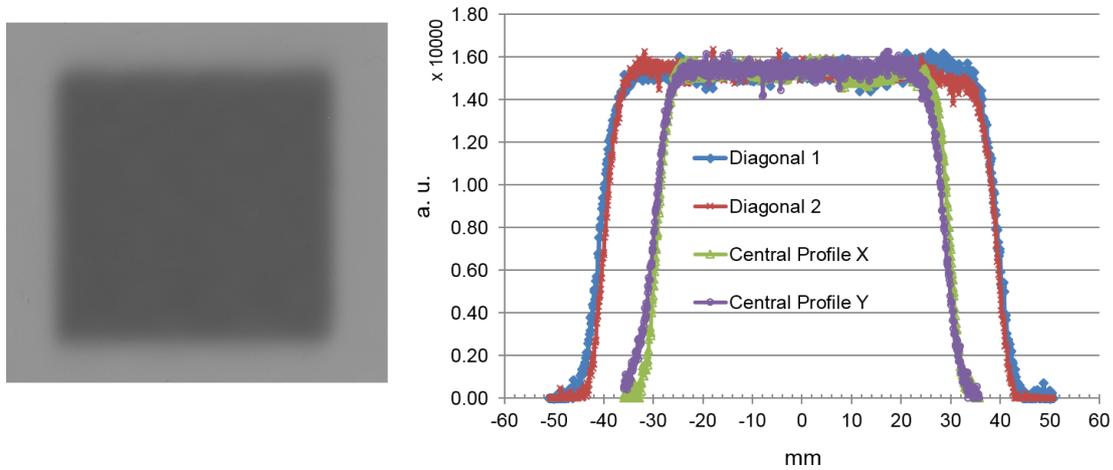

Fig. 12. Example of radiochromic film irradiated with a carbon ion beam: on the left a $60 \times 60$ mm$^2$ square field in air at the isocenter is shown while on the right the measured dose profiles along main axes (x and y) and diagonals are presented.

## V. DISCUSSION

A few clinical particle therapy centers in the world are providing a fluence modulated dose delivery and each center is equipped with its own dose delivery system tailored to the specifications of the accelerator and scanning magnets. These systems present some similarities with the system described here: the fluence is measured by a minimum of two ionization chambers [8, 9, 28, 29, 30] which provide the feedback to the system to stop the irradiation of each spot. All the systems are implementing some level of redundancy to ensure failure-safe operations and the ability to interrupt the irradiation with a maximum delay of a few hundreds of microseconds. The main differences originate from the delivery technique and from the details of the accelerator system. As an example, in the Paul Scherrer Institute [6, 28] and MD Anderson [8] delivery systems, the beam is stopped between each spot leading to a different flowchart of the delivery procedure [8, 29] and to a much less restrictive requirement on the fast control of the scanning magnets. Additionally, while the cyclotron-based systems are required to control several range shifters, the synchrotron systems communicate the proper beam energy to the accelerator control system.

The design of the DDS described in this paper started ten years ago with a prototype of the beam monitor chambers, fully tested [31] on the GSI clinical carbon ion beams.

The system reported here consists of two components: the beam monitors (BOX1 and BOX2) and the DDS control system. The established clinical protocols and the expected beam characteristics as fluence, shape and delivery time structure have been taken into account to define the system requirements. The technical specifications were driven by the clinical requirements but considering present technological limitations. The guidelines included the accuracy in beam steering and in flux measurements at a high sampling rate. Additionally, the design of the electronics was based on large scale integration to achieve a high frequency, dead-time free readout of hundreds of channels with uniform behavior thus reducing the need of a cross-calibration. The system was built to operate indefinitely, reading-out the detectors and receiving trigger events continuously. The control system, based on 5 FPGAs performing separate tasks, provides a processing speed and a data throughput well suitable for the required applications.

Particular care was devoted to patient safety, which lead to the design of a patient safety interlock system to timely detect potential hazards as they develop and to trigger the immediate stop of the irradiation.

The current system is well optimized for the treatment of static tumors. However, it has been designed to allow enough flexibility to implement tracking [32], gating or repainting techniques [33, 34, 35] to treat tumors in moving organs [36].

The ability of the DDS to accomplish a safe and accurate treatment was validated during the commissioning phase, some examples being reported in this paper. The high level of reliability and robustness has been verified in two years of activity with 100% of system up-time. In these years the DDS has rarely needed more than regular maintenance.

The careful analysis of the DDS data provided a thorough understanding of the performance of the system and stability in time. A preliminary study reported in [37] shows that in the first 10 months of clinical operation for over 95% of the spots the number of delivered particles differs by less than 1% from the prescription, with a good correspondence with the prescribed position. The impact on the dose distributions is currently under investigation.

The dosimetric accuracy was assessed during the patient specific Quality Assurance measurements showing dose deviations always lower than the acceptance threshold of 5%, with a mean deviation between TPS dose calculation and measurements lower than ±3% in 86% of the cases [15]. These results were found to be largely independent from the field shapes and in a wide range of number of particles per spot.

Several tests of the treatment recovery procedure were performed using dose measurements in phantom to verify the consistency of the dose for treatments where abnormal terminations have been forced (i.e. partial treatment delivery followed by treatment recovery managed by the OIS) leading to negligible deviations from the prescribed dose.

Daily checks of the conversion factors used to convert the number of particles into monitor counts are performed at CNAO for each treatment line and for protons and carbon ions separately. The results are found to be always within ± 2%, with a negligible dependence on the beam intensity.

## VI. CONCLUSIONS

The Dose Delivery System, specifically designed, developed and now in use in the clinical routine of CNAO, is described. It implements the modulated scanning technique with protons and carbon ions and, so far, has been used to treat more than 300 patients. This system has been proven to guide and control the therapeutic pencil beams with a high level of reliability, reaching performances well above clinical standards in terms of dose accuracy and stability. Examples of the system performance are reported in this work.

The interest in the technical details described in this paper is particularly relevant since the number of facilities treating patients with the modulated scanning technique is increasing worldwide.

## ACKNOWLEDGMENTS

The authors wish to acknowledge the designers and all the technical staff of CNAO, INFN and University of Torino who contributed to the development of the Dose Delivery System in the past ten years. A special thanks to Marco Lavagno for the strong help in the installation and commissioning of the system.

The authors wish to acknowledge Ugo Amaldi for the fundamental role in the birth of CNAO and Sandro Rossi for his leadership and advice during the entire span of the CNAO project.